Polymer-bonded magnets produced by laser powder bed fusion: Influence of powder morphology, filler fraction and energy input on the magnetic and mechanical properties

Kilian Schäfer[a], Tobias Braun[a], Stefan Riegg[a], Jens Musekamp[b], Oliver Gutfleisch[a]

[a] Functional Materials, Institute of Materials Science, Technical University of Darmstadt, 64287 Darmstadt, Germany

[b]Institute for Materials Technology (IfW), Technical University of Darmstadt, Grafenstraße 2, 64283 Darmstadt, Germany

Abstract

Bonded permanent magnets are key components in many energy conversion, sensor and actuator devices. These applications require high magnetic performance and freedom of shape. With additive manufacturing processes, for example laser powder bed fusion (LPBF), it is possible to produce bonded magnets with customized stray field distribution. Up to now, most studies use spherical powders as magnetic fillers due to their good flowability. Here, the behavior of large SmFeN platelets with a high aspect ratio as filler material and its influence on the arrangement and the resulting magnetic properties are examined in comparison to a spherical magnetic filler. The 3D distribution and orientation of the magnetic filler was studied by computed tomography and digital image analysis. The platelet-shaped particles align themselves perpendicular to the buildup direction during the process, which offers a new and cost-effective way of producing composites by LPBF with anisotropic structural and functional properties. The influence of LPBF parameters on the properties of the composites is investigated. Highest filling fractions are required for high magnetic remanence, however the powder itself limits this maximum due to particle shape and required minimal polymer fraction to form mechanically stable magnets. The coercivity decreases for higher filling fractions, which is attributed to increased rotation of insufficiently embedded magnetic particles in the matrix. It is discussed how filler morphology influences the observed change in coercivity since the rotation of spherical particles in comparison to platelet-shaped particles requires less energy. Our work shows the challenges and opportunities of large platelet shaped fillers used in LPBF for the production of anisotropic functional and structural composites.

1. Introduction:

Bonded permanent magnets are key components for many energy conversion devices including sensor and actuator applications [1]. These applications require high magnetic performance, customized stray field distribution and shapeability for complex geometries. Bonded magnets consist of magnetic particles within a polymer matrix and are conventionally processed via injection or compression molding [2]. The drawbacks of molding processes are the high price and limited shapes of tooling dies, which makes it economically only viable for large batch sizes and a reduced degree of freedom of shaping.

Additive manufacturing (AM) can overcome the limitations of conventional bonded magnets manufacturing techniques [3–5]. AM also brings the potential for manufacturing magnets with specific stray field distributions for sensor applications and improving them by the combination of soft and hard magnetic materials [6,7]. In addition, AM is ideal for net-shape production allowing the processing of magnets in a resource-efficient way due to the omission of additional subtractive machining. For lower volume parts the resource criticality of the rare-earth elements present in high-performance magnets can be addressed efficiently by AM [8].

To optimize the magnetic performance, the remanence $J_r$ (which scales to the square for the energy density $(BH)_{max}$) and the coercivity $H_c$ must be maximized [1]. A high remanence in bonded magnets can be achieved with a high magnetic filling fraction. However, the mechanical stability of polymer composites is reduced for high filling fractions [9] which is detrimental for applications with mechanical loads like motors or generators. Similar behavior can be observed in AM of bonded magnets [10]. Literature shows that the morphology of the filler has an influence on the magnetic, microstructural and mechanical properties of the resulting composites [11,12]. Fim et al. investigated the influence of laser parameters on the densification of bonded magnets [13]. Moreover, it was shown by Fliegl et al. that the porosity of parts produced with full metal laser powder bed fusion (LPBF) can vary across the powder bed depending on the laser incident angle [14].

The relations between process parameters, filler fraction and morphology on the magnetic and mechanical performance of bonded magnets produced by AM processes are the subject of current research but are yet to be fully understood [3,12,13,15,16]. Due to its high flowability, which influences the processability of the magnetic filler in LPBF, and its easy availability, the focus of previous research on AM of bonded magnets has mostly been on the utilization of spherical, gas-atomized powders with particles sizes of around 20 – 50 µm [3]. The behavior of large $Sm_2Fe_{17}N_3$ platelets with a high aspect ratio as filler material, its influence on the distribution and arrangement, as well as the resulting magnetic mechanical properties has not been studied in detail yet. Up to now, similar studies used significantly smaller particle sizes for the AM processes with platelet-shaped powders [12,16]. The $Sm_2Fe_{17}N_3$ powder used in this manuscript has median particle size of 91 µm.

The aim of this work is to investigate the trade-off between the magnetic and mechanical properties of polymer composites produced by LPBF while studying in detail the differences caused by magnetic fillers with significantly different morphologies. In this study, spherical NdFeB and significantly larger platelet-shaped SmFeN particles with a high aspect ratio were utilized as filler materials to investigate the potential of non-spherical particles in AM of bonded magnets. The influence of the laser energy density and temperature distribution across the build volume on the magnetic and mechanical properties is studied and compared to the performance of pure polymer polyamide 12 (PA12). Furthermore, the impact of an increasing filler fraction on the microstructure, magnetic and mechanical properties for both types of morphologies is explored. This paper also presents the potential of the different used morphologies for an *in situ* alignment process of the magnetic filler. This alignment effect can offer new and cost-effective ways of producing anisotropic bonded magnets by LPBF, which show higher magnetic performance in comparison to isotropic bonded magnets. The alignment process can also be transferred to AM of other functional composites produced by LPBF to specifically tune properties like electrical conductivity.

2. Experimental:

Bonded magnets were produced using PA 12 from *Sintratec* (*Sintratec*, *Switzerland*), in combination with commercially available magnetically isotropic powders: MQP-S (MQP-S-11-9) which is $Nd_2Fe_{14}B$ based from *Neo Magnequench* (*Singapore*) and $Sm_2Fe_{17}N_3$ (*Daido Steel*, *Japan*). The powders were used in their as received, thermally demagnetized state from the manufacturer. PA12 provides a good combination of flowability and good mechanical properties which qualifies it very well for the processing of highly filled composites. Therefore, it is used as the matrix material for the LPBF process in this study. The morphology of the magnetic powders is presented in Figure 1. The Scanning electron microscopy (SEM) images show that the MQP-S consist of rounds particles, as produced by atomization, with an average particle size of 35-55 µm [17]. On the other hand, the $Sm_2Fe_{17}N_3$, which is produced by melt spinning, consist of larger platelet shaped particles with an average size of 91 µm [18]. Latter will be further referred to as SmFeN. The full hysteresis loops and initial magnetization curves of the two magnetic powders, measured in a PPMS at 300K up to 4T for fixed powders, are presented in Figure 2. Both powders are so called exchange-coupled magnetic materials [18,19] which can be seen by the ratio between remanence polarization ($J_r$) and saturation ($J_s$). This ratio is larger than 50% of $J_s$. For isotropic magnetic materials remanence is half of the saturation. In exchange coupled permanent magnets, the coupling between a hard and soft magnetic phase leads to the stabilization of the soft magnetic phase by the high coercive phase during the demagnetization process; for this the soft phase should be well-below 100nm in size [19]. Due to the hindered switching of the soft magnetic phase, an increase of the remanence can be observed [18]. The properties of the feedstock materials as given by the manufacturer are presented in Table 1.

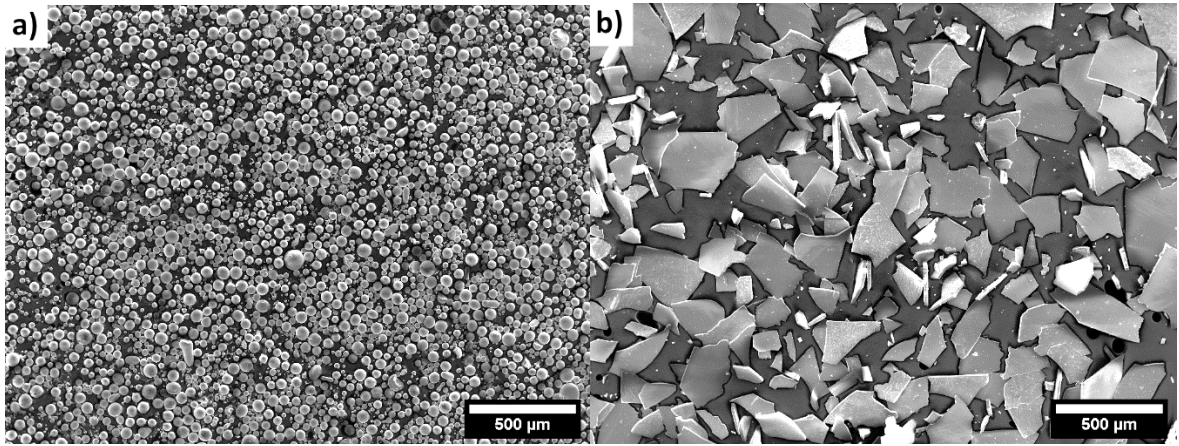

Figure 1: SE-SEM images of the initial powders: a) spherical MQP-S on Nd-Fe-B basis and b) platelet-shaped $Sm_2Fe_{17}N_3$.

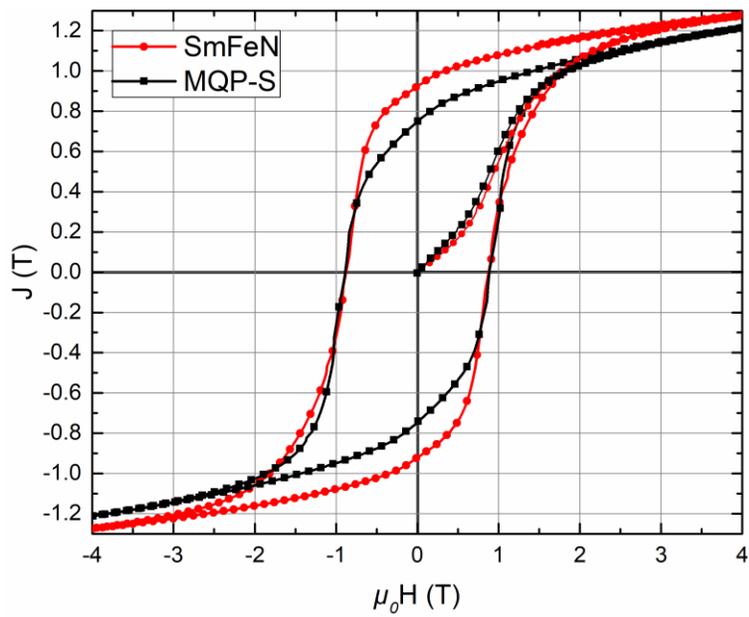

Figure 2: Full hysteresis loops of the two magnetic powders MQP-S and SmFeN including the initial magnetization curve.

Table 1: Properties of the feedstock powders as listed by the manufacturer.

| Powder | Material | Morphology | Particle size $D_{50}$ (µm) | Theoretical density (g/cm$^3$) | Remanence $J_r$ (T) | Coercivity $\mu_0H_c$ (T) |
|---|---|---|---|---|---|---|
| PA12 Sintratec [20] | Matrix | Spherical | 60 | 1.1 | | |
| MQP-S Neo-Magnequench [17] | Filler | Spherical | 35-55 | 7.43 | 0.75 | 0.84 – 0.94 |
| SmFeN Daido steel [21] | Filler | Platelet | 91 | 7.66 | 0.92 | 0.85 – 1.0 |

The LPBF device used in this work is the commercially available "*Sintratec Kit*" (*Sintratec, Switzerland*) equipped with a 2.3 W blue diode (λ = 445 nm) laser. The manufacturing parameters used are listed in Table 2. As the surface temperatures are significantly affecting the macroscopic features (e.g. curling of the manufactured layer) of the samples, the lowest possible surface temperatures where no curling occurred where chosen for the experiments. To minimize the unused build volume of the manufacturing, a self-designed insert was used for the reduction of the building surface from 13 x 13 cm$^2$ to 5.9 x 9.4 cm$^2$.

Table 2: LPBF parameters used with the LPBF device Sintratec Kit

| Layer height | 100 µm |
|---|---|
| Hatch distance | 0.3 mm |
| Laser scanning speed | 300 mm/s |
| Focus diameter | 0.25 mm |
| Surface temperature | 162.5 °C (for composites) |
| | 167.5 °C (for pure PA 12) |
| Chamber temperature | 145 °C |

Samples were produced at different positions with respect to the center of the powder bed as shown in Figure 3. The positions are designated with the letters *A, B, C, D, E, F, G, H, I, J, K, L* and *M*. Since the positions are arranged in a circular manner, the following positions have the same distance from the center: *F, G, H* and *I* (1.41 cm), *B, C, D* and *E* (2.00 cm) and *J, K, L* and *M* (2.82 cm), respectively. The position of the IR lamps (for surface temperature), the resistive heater (for chamber temperature) and the laser are shown schematically. The temperature of the powder surface is measured close to position *A* with an infrared pyrometer.

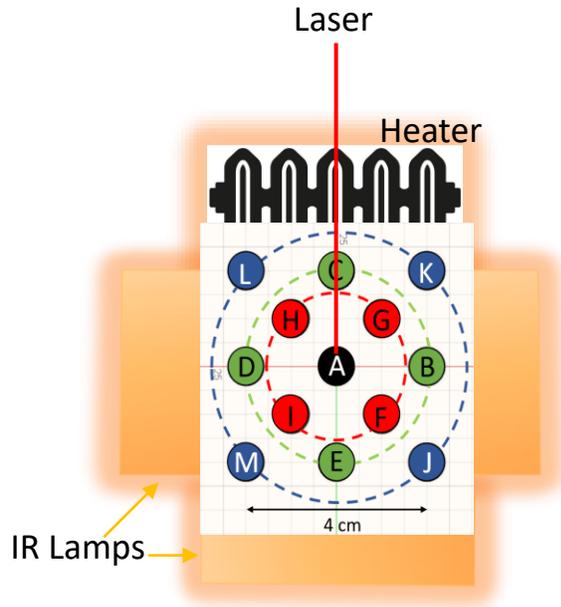

*Figure 3: Schematic image of the sample positions and setup of the heat sources in the Sintratec KIT (Top view).*

For the printed samples a cylindrical geometry with the diameter of 8 mm and height of 16 mm was selected as it is described in the norm for compression tests EN ISO 604:2003 [22]. According to this norm that the ratio of diameter to height should be ≥ 0.4. Since geometrical deviations from the nominal CAD model are typically found for additive manufactured parts, a slightly bigger ratio of diameter to height of 0.5 was used.

To investigate the influence of the magnetic filler fraction on the properties of bonded magnets, several feedstocks were produced as presented in Table 3. The feedstocks of the composites were produced by mixing in a shaking mixer *TURBULA®* (*WAB-GROUP*, Switzerland) for 30 min. The PA 12 was used as received from the manufacturer (for properties see Table 1). To investigate the influence of the printing position, samples of pure PA 12 and composites with a nominal filler fraction of 80 wt.% were produced on all positions as indicated in Table 3. For the investigation of the filler fraction influence, only the center position (Position *A*) is discussed by means of the density and magnetic properties. For the analysis of the mechanical properties, the inner four positions (*F, G, H* and *I*) were examined as well.

Table 3: Nominal feedstock composition and the printing positions for the cylindrical samples

| Feedstock | filler content [wt.%] | Positions |
|---|---|---|
| Pure PA12 | 0 | A,B,C,D,E,F,G,H,I,J,K,L,M |
| MQP-S 70 | 70 | A,F,G,H,I |
| MQP-S 80 | 80 | A,B,C,D,E,F,G,H,I,J,K,L,M |
| MQP-S 90 | 90 | A,F,G,H,I |
| MQP-S 95 | 95 | A,F,G,H,I |
| MQP-S 97.5 | 97.5 | A,F,G,H,I |
| SmFeN 70 | 70 | A,F,G,H,I |
| SmFeN 80 | 80 | A,B,C,D,E,F,G,H,I,J,K,L,M |
| SmFeN 90 | 90 | A,F,G,H,I |
| SmFeN 95 | 95 | A,F,G,H,I |
| SmFeN 97.5 | 97.5 | A,F,G,H,I |

A *VEGA3* scanning electron microscope (*TESCAN, Czech Republic*) equipped with a tungsten cathode (0.2 to 30 kV) electron source was used in Secondary Electron (SE) mode to study the particle morphology of the initial powders and Backscattered Electron (BSE) mode to analyze the LPBF composite samples.

The geometric densities $\rho_{geo}$ of the samples were determined according to Equation 1 by determining the radius and height with a caliper and measuring the weight:

$$\rho_{geo} = m/V = m/{r^2 \cdot \pi \cdot h} \qquad \text{Eq 1}$$

where, *m* is the mass of the sample, *V* is the volume for cylindrical shapes (radius *r*, height *h*). The measurement uncertainty was estimated with a Gaussian error propagation calculation according to:

$$\Delta\rho_{geo} = \sqrt{(\partial\rho_{geo}/\partial m \cdot \Delta m)^2 + (\partial\rho_{geo}/\partial h \cdot \Delta h)^2 + (\partial\rho_{geo}/\partial r \cdot \Delta r)^2} \qquad \text{Eq 2}$$

$$\Delta \rho_{geo} = \sqrt{(1/_{r^2 \cdot \pi \cdot h} \cdot \Delta m)^2 + (-m/_{r^2 \cdot \pi \cdot h^2} \cdot \Delta h)^2 + (-2m/_{r^3 \cdot \pi \cdot h} \cdot \Delta r)^2} \qquad \text{Eq 3}$$

Here *Δm*, *Δr*, and *Δh* are the measurement errors of the mass, the radius, and the height. The mass of the cylinders was measured with an analytical scale with a precision of 0.001 g. The radius and the height were measured with a caliper with a precision of 0.002 cm. The measurement error of the density results to approximately *Δρ*$_{Bulk}$ = ±0.025 g/cm³ for all samples. For comparison, the theoretical density of the composites can be calculated by:

$$\rho_{Theo} = \phi_f \cdot \rho_f + (1 - \phi_f) \cdot \rho_m \qquad \text{Eq 4}$$

where $\phi_f$ is the filler fraction, $\rho_f$ the density of the filler and $\rho_m$ the density of the matrix. The full hysteresis loops of the initial magnetic powders were measured with a PPMS-VSM (Quantum Design PPMS-14) at room temperature under an applied magnetic field of up to 4 T. The measured hysteresis loops were corrected with a demagnetization factor *N* = ⅓ due to the separated isotropic orientation of the measured particles.

The J(H) demagnetization curves of the composites were measured at room temperature under external magnetic fields up to 6.8 T with a pulsed field magnetometer (Metis Instruments, Belgium) to determine the remanent polarization *J*$_r$. Since the measurement is performed in an open magnetic circuit, the demagnetization field must be considered. For the demagnetization correction, a value of *N* = ⅓ for the MQP-S composites is selected considering the spherical morphology of the magnetic particles. For the platelet shaped particles, a value of *N* = 0.71 was used. The platelets are aligned perpendicular to the build-up and measurement direction, as shown in Figure 7. The demagnetization factor is approximated based on a platelet geometry with quadratic shape where the short dimension is parallel to the measurement direction. Considering the micrographs and the literature [18] of the magnetic powders the average dimensions of 15 µm x 83 µm x 83µm were determined and these values are used for the estimation of the demagnetization factors $N$ for these samples. With this approximated mean geometry, the demagnetization factor was calculated according to reference [23] with the measurement direction parallel to the shortest measure.

To analyze the mechanical properties, compression tests were performed at room temperature using the Instron Universal Testing System (model 5900) with a 30 kN load cell. The cylindrical specimens were tested at a constant traverse velocity of 8 mm/min to generate a velocity rate of 0.5 1/min of the sample length to fulfill the compression test norm as described in [22]. The displacement of the compression plates was used to evaluate the sample strain.

The investigations of the actual filler content of the printed samples from position A were done by thermogravimetric analysis (TGA) with small parts of the samples after the compression tests. The TGA was performed with a *STA 409 CD Simultaneous thermal analyzer (Netzsch, Germany)* and heating rate of 10 °C/min under nitrogen atmosphere. The temperature range of the measurement was selected to be 30 – 600 °C.

The microstructure of the composite magnets was additionally investigated by computed tomography (CT) images with a *v|tome|x 240d (General Electric, USA)* with a nanofocus tube. An acceleration voltage of 120 kV, a current of 80 µA and 1000 ms exposure time resulted in a resolution of 7.7 µm voxel edge length. To shift the intensity of the X-rays towards softer radiation, two filters were used: 0.5 mm Sn and 0.5 mm Cu. After reconstruction, the specimens were registered and analyzed with *VGStudioMAX 3.0*.

The alignment of the SmFeN magnetic filler in the printed part was analyzed on either SEM images or cross-sections from the CT-images. The images were cropped around a center portion of 300 by 300 pixel² (>80 %) to exclude edge effects of the CT-scan and provide an even particle distribution for the analysis. This orientation analysis was done using *Fiji (ImageJ)* v1.52 [24] utilizing the *OrientationJ* plugin v2.0.5 which analyses the local gradient of the structure tensor for the images [25]. The analysis was performed with the accuracy of 1°, a local window $\sigma$ of 3 pixels and the cubic spline gradient. The orientation of the build-up direction was vertical for all analyzed images, which corresponds to an angle of 90° in the distribution. To provide statistical significance the orientation analysis was performed on at least 35 cross sections generated from CT-Scans for each filler fraction. The cross-sections are created from multiple sides of the scan but always parallel to the build-up direction. For each image, the orientation analysis was performed, the relative frequencies of each angle were calculated and the results were combined to one average orientation distribution for each filler fraction. Positive (0.5° to 90°) and negative (-0.5° to -90°) alignment angles were summed, as only the absolute value of the platelet tilt is of relevance.

3. Results and Discussion

   3.1. Influence of laser energy density and temperature distribution in the powder bed on density and mechanical properties:

The laser energy density transferred to the powder and the powder bed temperature have a significant influence on the resulting mechanical properties of the printed parts [26]. It is not surprising, that these quantities vary over the position of the printed parts during the LPBF process [27,28]. Therefore, it is necessary the special distribution of the properties in the powder bed to increase reproducibility of the part quality. A key parameter for composite parts produced with LPBF [29] is the density. The geometrical density, which is determined according to Eq. 1, as function of their position in the powder bed is shown in Figure 4. The density distribution for samples consisting of 80 wt.% MQP-S and PA12 shows a radial symmetry around the center. The density of the samples varies from 2.98 g/cm$^3$ at the center of the printing area to 2.23 g/cm$^3$ at the corners of the area. The theoretical maximum density for this filler fraction is 3.27 g/cm$^3$. Porosity in the printed parts is responsible for the difference of the theoretical density and the measured values. The geometric density of the samples from the K and L position is higher than the density of the *M* and *J* positions.

The origin of the density distribution is the mutual interference of the varying energy density of the laser and the heat distribution in the powder bed resulting from the resistive heater and the three IR lamps as shown schematically in Figure 3. Since the movement of the laser spot is obtained by deflecting the laser beam through rotary mirrors actuated by galvanometers, the energy density of the laser is not homogenously distributed through the powder bed [30]. In addition, the other heat sources in the printer: the resistive heater behind the powder bed and the three infrared lamps also influence the heat distribution. Therefore, the temperature in the powder bed is higher closer to the resistive heater (Positions *K* and *L*), as the overall temperature of powder bed is not monitored across the whole powder bed, but rather at the center of it. This explains the higher density of the *K* and *L* positions in comparison to the *M* and *J* positions.

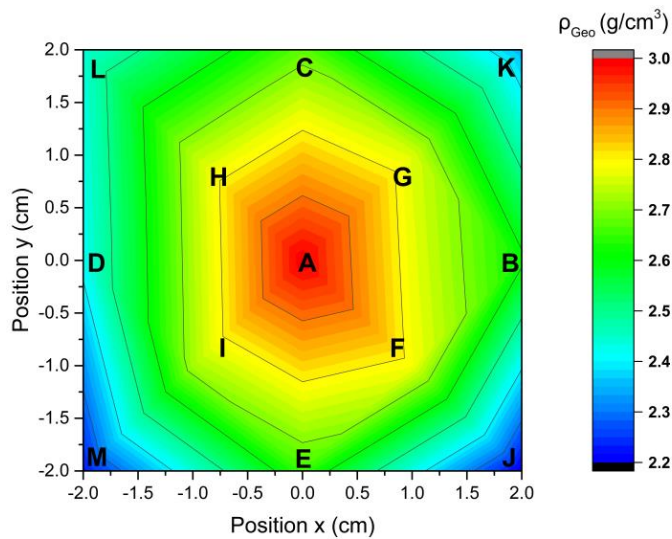

*Figure 4: Geometrical density distribution in the powder bed for the 80wt.% MQP-S composite. The density is radially decreasing with increasing distance from the center of the powder bed.*

Figure 5 depicts the dependence of density (Figure 5 a/c) and compressive stress (Figure 5 b) on the distance to the center. Figure 5 a) shows the decrease of the density of the SmFeN composites and pure

PA12 in accordance with increasing distance from the center. The variation of the properties of the samples is indicated by their standard derivation. The average value for one distance is the mean value of the four positions within this distance. The density of the outer positions (*J,K,L* and *M*) is, in comparison to the center position, reduced by 22 %, 17 % and 13 % for the SmFeN composites, MQP-S composites and pure PA 12 samples respectively as shown in Figure 5 c. This is in agreement with the power distribution described above, which results in insufficient consolidation of the polymer particles, which corresponds to porosity formation.

The compressive stress at yield, presented in Figure 5 b) decreases with increasing distance towards the center as well. The porosity, crystallinity and the degree of remolten to unmolten polymer (degree of particle melt) have an influence on the mechanical properties of PA12 processed with LPBF [31,32]. High porosity is often accompanied by poor particle fusion [32], which results in lower mechanical stability of samples with lower density [29]. The other two main factors for the mechanical stability, the crystallinity and the degree of particle melt are also influenced by the amount of thermal energy transferred to the polymer.

Figure 5 c) shows that the relative difference in the density is larger for composite materials. This is an indication that the process window for composite materials is smaller in comparison to pure polymers. That means that the effects of inhomogeneous laser energy distribution and temperature must be verified and considered especially when composite materials are being processed with LPBF.

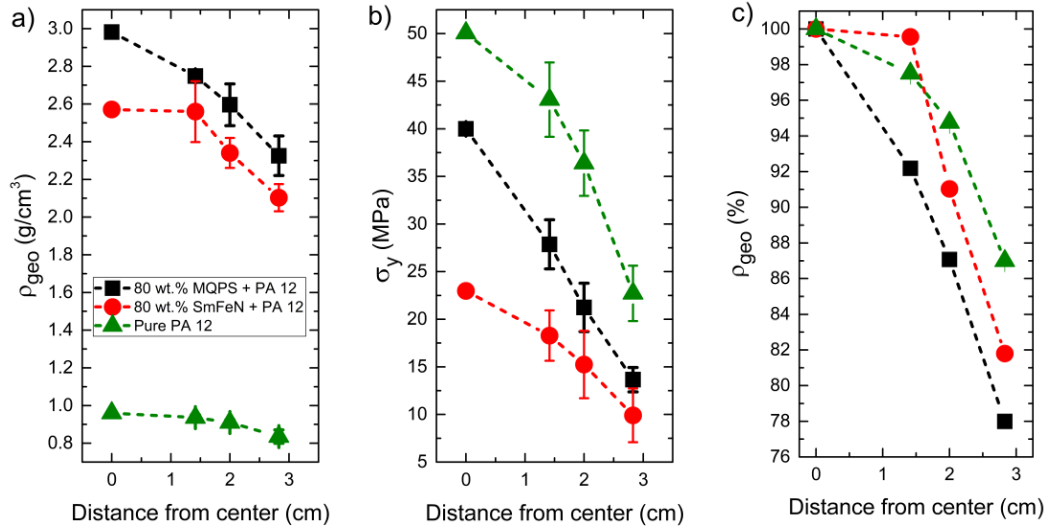

*Figure 5: a) Geometric density, b) compressive stress at yield as function of the distance from the center of the powder bed and c) showing the relative change of the geometric density as function of the distance from center. A measurement uncertainty was determined for all data points but might not be visible in the graph due to the size.*

## 3.2. Influence of filler fraction and orientation of MQPS and SmFeN composites

### 3.2.1 Determination of filler fraction

The nominal filling fraction as mixed wt.% is verified by TGA measurements. The results for the MQP-S composites are shown in Figure 6 a) and the SmFeN composites in Figure 6 b) as mass percentages of the composites as a function of the temperature. A decrease starting at around 375 °C can be observed, which is caused by the decomposition of the PA12. The onset of this decomposition increases with higher filling fraction. This is known as the shielding effect and can be explained with faster heat absorption of the fillers in comparison to the polymer [33,34].

*Table 4: Nominal and measured filler weight percentage for the MQP-S and the SmFeN composites.*

| Nominal filler [wt.%] | Measured MQP-S [wt.%] | Measured SmFeN [wt.%] |
|---|---|---|
| 70 | 70.0 | 73.1 |
| 80 | 83.9 | 80.2 |
| 90 | 91.8 | 89.7 |
| 95 | 96.0 | 94.9 |
| 97.5 | 97.8 | 97.6 |

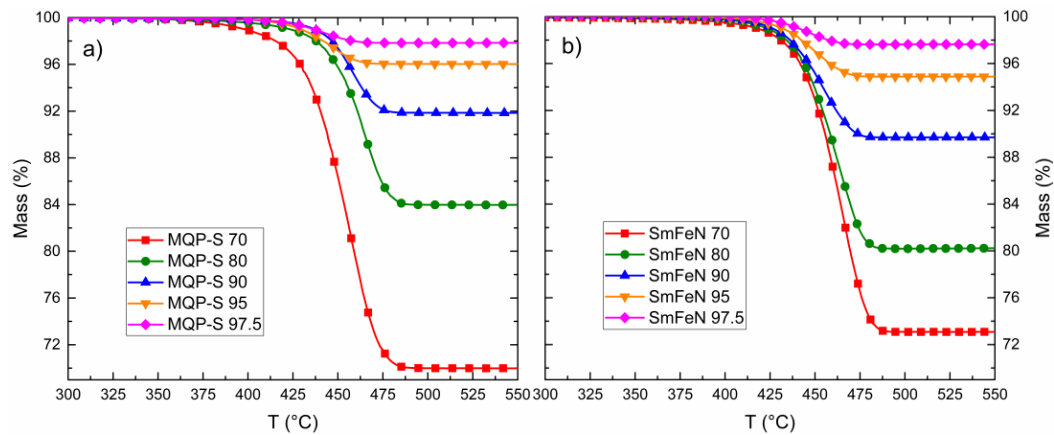

*Figure 6: TGA of a) MQP-S composites and b) SmFeN composites with different target weight fractions of magnetic filler. Due to the decomposition of the polymer at high temperatures, the remaining mass is the filler fraction of the composite.*

The filler fractions determined by the TGA analysis are given in Table 4 and confirm that the actual filler fractions are close to the nominal values. The largest deviation is around 3.9 wt. % for the MQP-S 80 composite. Deviations from the nominal composition can occur during the production of the samples by LPBF. The mixing process and loading of the powder mixture into the printer can have an influence in the actual filler fraction, as the powder mixture can separate due to the significant differences in density between the PA12 matrix and magnetic filler materials.

*3.2.2 Influence of the powder morphology on the distribution and orientation of the filler*

Figure 7 depicts SEM images of the MQP-S 70 (a) and the SmFeN 70 (b) samples. The built-up direction is vertical and marked in Figure 7 with a red arrow for better visualization. As can be seen, the MQP-S particles are dispersed homogeneously over the sample. In contrast to this, the SmFeN particles are mostly aligned perpendicular towards the build-up direction. This can be explained by two reasons. First, it is more likely that platelet-shaped particles align themselves horizontally in the powder bed due to gravitational forces. A similar behavior was also reported for smaller particles in polyurethane composites [35]. Erb et al. predicted and verified experimentally the alignment of platelet shaped particles due to gravitational forces, if the particles are larger than a few micrometers [35]. Secondly, the sweeping of the blade during the powder deposition process could help to align the particles since the sweep movement is parallel to the long side of the platelets.

This automatic self-alignment of the platelets during the build-up offers an efficient way to produce textured anisotropic magnetic composites without the necessity for an additional external stimulus if a material system with a magnetic easy axis perpendicular to the platelet plane is used. As this effect does not depend on any applied field during the printing process or any subsequent processing steps, it has the potential to make the processing of anisotropic composite magnets by LPBF easier and more cost effective.

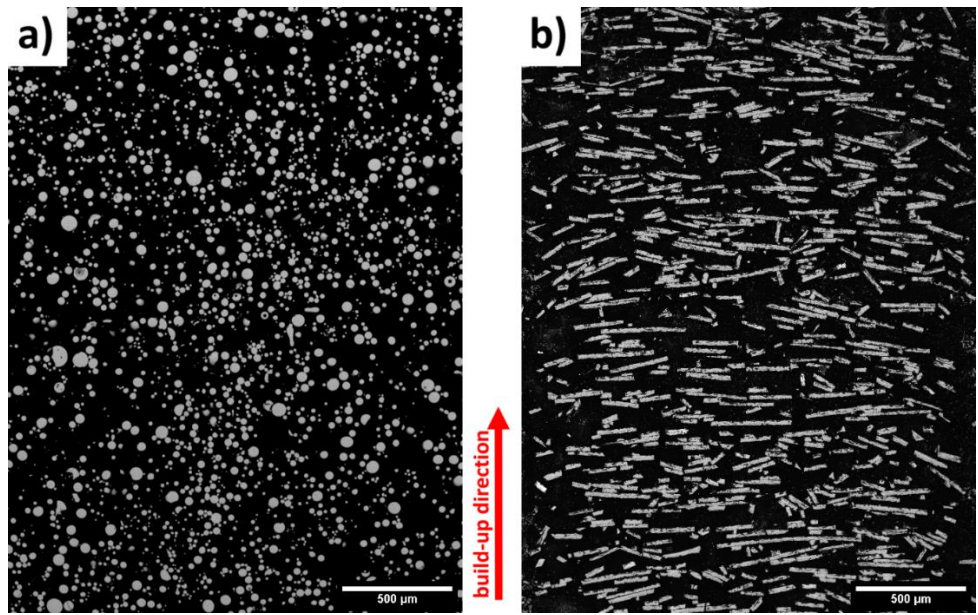

*Figure 7: BSE-SEM images of composites produced with LPBF: a) MQP-S and b) SmFeN composites. The vertical build-up direction.*

A quantitative analysis of the alignment for the platelet shaped particles is carried out according to the earlier described procedure utilizing *ImageJ*. An example for the orientation analysis is shown in Figure 8 for the SmFeN 70 sample with the build-up direction arranged vertically. Misoriented plates show a significant deviation from the perfect alignment angle of 0°, parallel to the build-up plane, and smaller particles tend to align less within the printing layer then larger platelets. The resulting distribution shows a slight increase around +75° to +80°, which corresponds to the red colored smaller platelets seen in the mapping. The increase at +40° is caused by the platelets shown in dark blue, mainly located at the left edge of the mapping. Since the goal of most magnetic materials is a potentially high $J_r$, the influence of an increasing amount of magnetic filler to reach this goal is of high interest. The shown distribution analysis allows a quantitative comparison of the alignment for different filler fractions. It also suggests that smaller platelet shaped particles can result in a worse alignment during the printing process, however, further research on investigating the influence of the particles size distribution has to be performed to prove this.

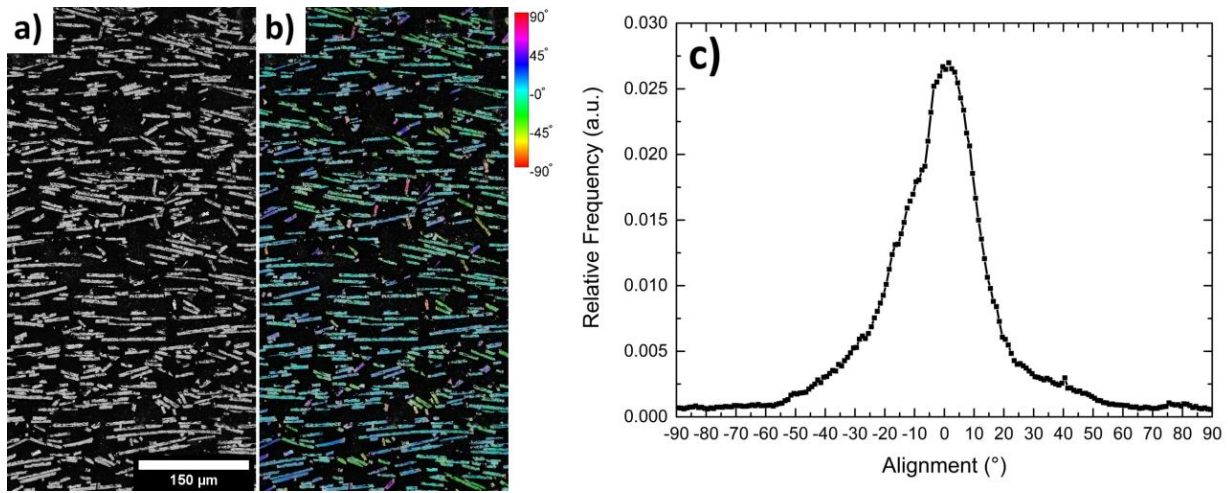

*Figure 8: SEM-BSE image (a) of composite used in mapping of the optical orientation analysis (b) for a SmFeN 70 sample with vertical build-up direction done with ImageJ and corresponding angular distribution (c).*

Figure 9 displays the averaged relative frequencies of the orientation analysis for each SmFeN sample. For an increase of the filler fraction from 70 to 95 wt.% a significant broadening of the angular distribution can be observed. The median of the distribution of the misalignment angle from the build-up plane shifts from 7.5° to 10.5°. This broadening of the distribution with the filler content is attributed to the increasing probability of platelets in direct contact pushing each other out of the horizontal build-up plane during the application of the powder layers. This allows lower filler fractions to reach a higher alignment. A further increase of the filler fraction to 97.5 wt.% shows a median and distribution similar to that of 90 wt.%. A possible explanation for this is a forced alignment of the platelets during the build-up to form a higher packing density layer. With such high filler fractions, a higher amount of random oriented plates is less favorable during the powder layer application, since the packing density should become the critical factor determining the density of the composite and therefore influencing the texture. To verify this theory, the processing of even higher filler fractions and additional investigations become necessary.

Without the necessity for additional process steps or alterations, the discussed self-alignment effect for platelet shaped particles allows for a significant texture, depending on the amount of filler particles used. Considering the pursuit of a high $J_r$, a high fraction (above 95 wt.%) of large, platelet shaped magnetic filler can be favorable for both alignment and overall achievable magnetization if other properties like $H_c$ or mechanical properties are within the desired range. This effect could also be transferred to the processing of other structural or functional composites produced by LPBF to tune the mechanical or functional properties.

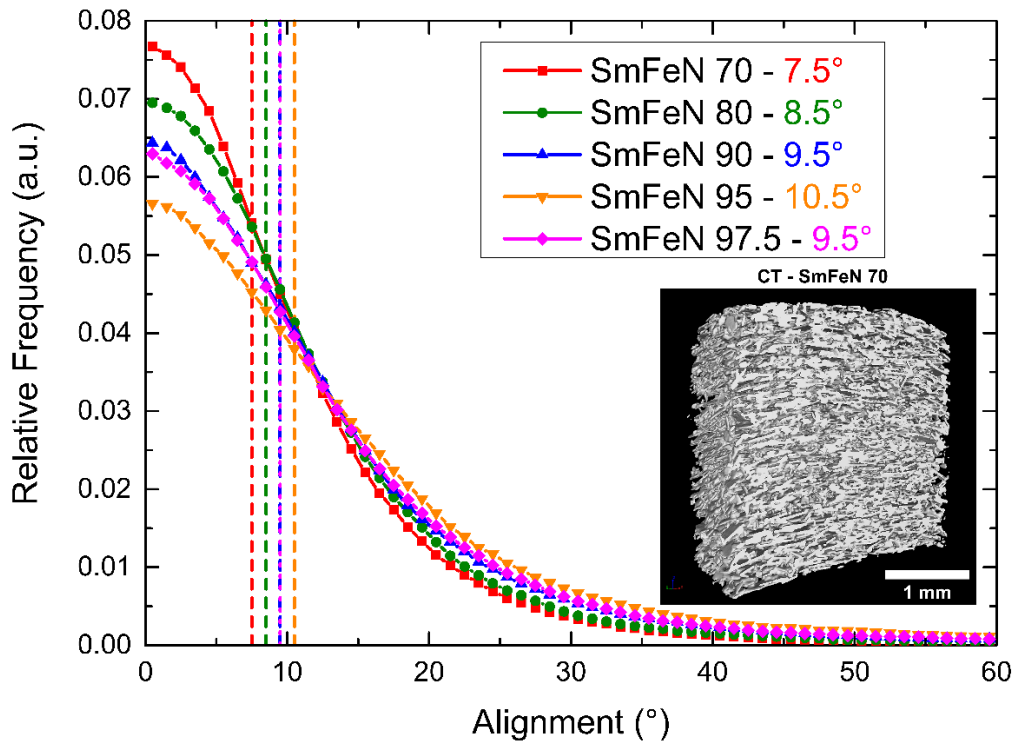

*Figure 9: Averaged relative frequencies of the optical orientation analysis for the self-alignment effect of SmFeN composites of different filler fractions with alignment perpendicular to the build-up plane at 0° evaluated from CT scans. The dashed lines mark the median alignment angles and show a broadening of the angular distribution for higher filler fractions with a partial recovery for 97.5 wt.%. The numerical values of the median alignment are given in the legend of the graph. The inset image shows the cross-section of the CT scan for the SmFeN 70 sample for better visualization.*

*3.2.3 Geometric density*

Taking into account the previously discussed influence of the heat distribution and the change of laser energy density across the printed area, further analysis of properties was performed on the central position (position *A*) for all compositions. For the mechanical evaluation, the samples from the center positions *A, F, G, H* and *I* are investigated to increase the statistical significance of the data. The filler fractions are the ones determined with TGA and can be found in Table 4.

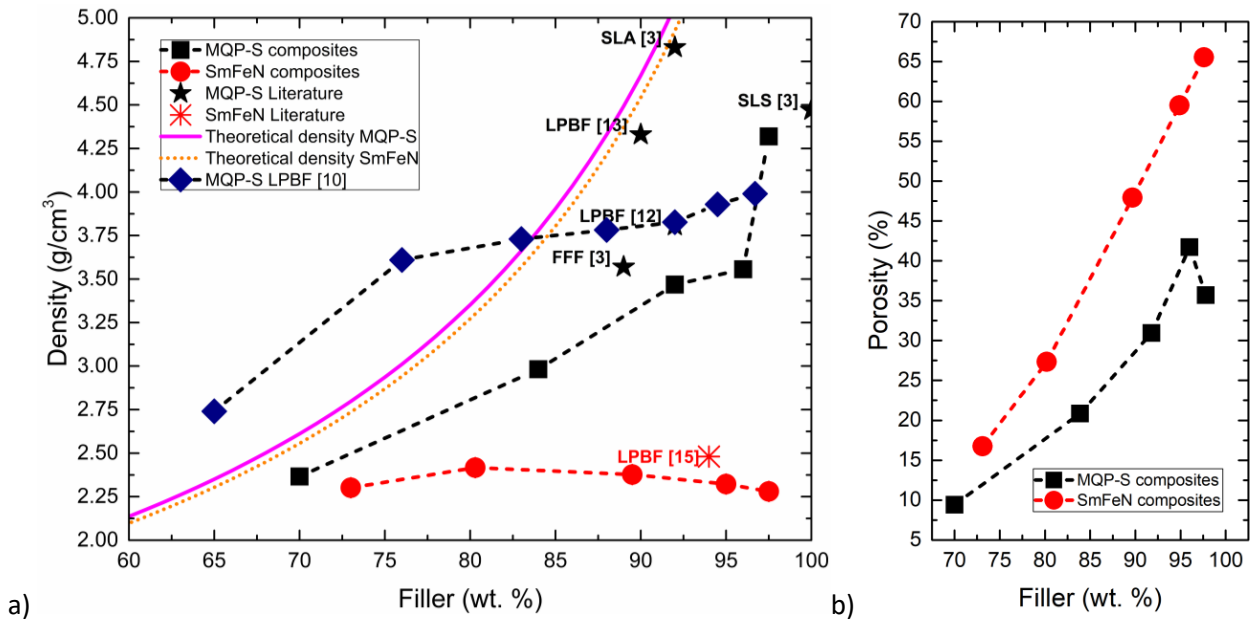

*Figure 10 a): The change of geometric densities of the produced samples as function of the filler wt.%. The solid lines represent the upper theoretical limits and symbols indicate the results obtained in this work and reported in literature. b) Porosities of the composites as function of the filler wt.%.*

Figure 10 a) shows the geometric density of MQP-S and SmFeN composites as function of their filler wt.%, in comparison to literature values and the theoretical density, which was calculated with Eq. 4. The values from the literature are given in Table 5. The porosity values of the samples which were calculated by the relation of the measured density values to the theoretical density are presented in Figure 10 b). The geometrical density of the MQP-S composites increases with increasing filler fraction due to the significantly higher density of the MQP-S powder in comparison to the P12 polymer. The geometric density of the SmFeN composites at first increases from the 70 to 80 wt.% samples and then decreases with increasing filler content. The density of the composite is limited by the highest achievable packing density of the magnetic particles since they are not molten during the LPBF process. For spheres with the same size, the highest achievable packing density is 74 vol. %. For a bimodal mixture of spherical particles with large and small particles, a maximum density of 86 vol.% can be reached if the mixture contains 70 % large particles [36,37]. For the MQP-S particles, the highest achievable packing density is somewhere between 74 and 86 vol.% since it consists of spheres of different sizes as shown in Figure 1. A vol. % of 74 - 86 relates to a wt.% of 95 – 97 within a PA12 matrix. For the platelet-shaped SmFeN particles, the packing density could in theory be significantly higher than for the spherical MQP-S particles, if they align mostly parallel due to their high aspect ratio. The given values show that the maximum packing density should only be the limiting factor for an increasing density, if a very high amount of magnetic filler, above 95 wt.%, is utilized. Therefore, the deviations from the theoretical density, occur below the limit predicted by the highest achievable packing density. This can also be shown with the increase of the porosity for the composites with increasing filler fraction. For example, the 70 wt.% MQP-S sample has a geometric density of 2.36 g/cm$^3$ but the theoretical density is 2.61 g/cm$^3$ (Equation 4) which is equivalent to 9.41 % porosity. Several explanations are possible for this behavior. Particle rearrangement in a liquid phase is seen as an important process for the complete densification of a composite [38]. If not enough liquid phase is present or the time is too short for particle rearrangement, the formation of pores is facilitated [32]. If a high filling fraction is present in the powder bed, a high fraction of the laser energy is absorbed by the filler particles, which have a higher heat capacity, and not by the polymer. Therefore, less energy is available for the melting of the polymer. A low melt viscosity is an additional beneficial factor for consolidation of the polymer [39]. A high filler fraction significantly increases the melt viscosity [40]. This can also be responsible for the formation of porosity of the samples with a high filling fraction. The results suggest that the consolidation of MQP-S and PA 12 with the used printing parameters is insufficient, which leads to the formation of porosity. However, further optimization of the printing parameters can partially solve these problems and result in higher densities. Fim et al. achieved a geometrical density close to the theoretical density of a 90 wt.% MQP-S composite by optimization of the printing parameters [13].

A similar LPBF process using MQP-S composites by Mapley et al. results in a comparable density [12] to the presented results in this study. However, in another recent study by Mapley et al. [10] the actual filler content in the composites is not verified, which gives some uncertainty for the reported densities. A Fused deposition modeling (FDM) process from Huber et al. results in a slightly lower density, whereas a stereolithography (SLA) process results in a high density of 4.83 g/cm$^3$, which is in accordance with the theoretical density. The reason for the high density of the SLA sample is that the magnetic powder is mixed with a liquid photo-reactive feedstock utilizing a centrifuge [3]. Therefore, the liquid feedstock embedded all particles without pores. The metallic SLS sample has a density close to the tap density of the MQP-S powder [3]. This means that the MQP-S powder is not completely molten.

The geometrical densities of the SmFeN composites are lower than their theoretical density and the values of the MQP-S composites for all wt.%. Since the densities of the metallic powders itself are very similar (MQPS: 7.43 g/cm$^3$, SmFeN: 7.66 g/cm$^3$) this direct comparison is reasonable. It suggests that the platelet morphology of the SmFeN particles can increase porosity during the LPBF process, consistent with Mapley et. al [12], where the density of samples with the same wt.% for spherical particles was 18 % higher in comparison to flake shaped fillers. As an example, for 80 wt.%, the bulk density of the MQPS particles is 35 % higher in comparison to the SmFeN counterpart. This is also reflected in an increase of the porosity for the SmFeN based composites. One reason for this is the higher achieved packing density of the spherical powder. In general, the more spherical the particles are, the better the densification during the LPBF process [41], which is beneficial to prevent porosity formation [32]. The highest achievable packing density of the SmFeN platelets is difficult to estimate due to their alignment as shown in Figure 7 b). For the LPBF study with ball milled SmFeN powder by Engerroff et al., bulk densities of 2.30 to 2.48 g/cm$^3$ for 94 wt. % SmFeN were reported [14]. The higher density reached is caused by the additional milling of the SmFeN platelets which result in a significantly reduced particle size of 35 - 55 µm and increased roundness.

### 3.2.4 Magnetic properties

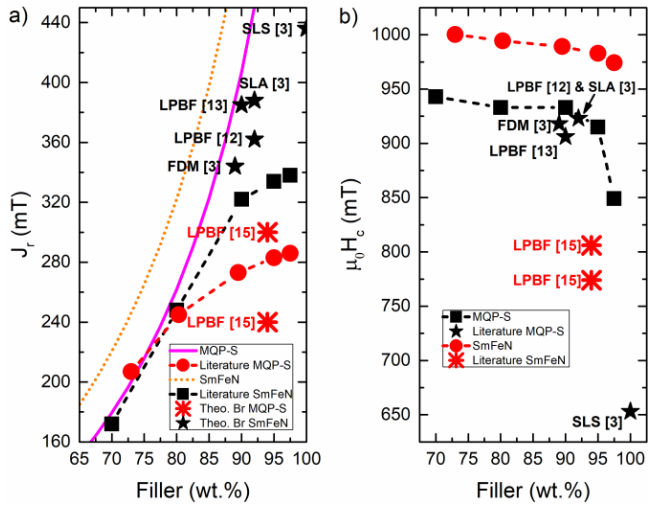

*Figure 11: a) Remanence as function of the filler wt.% in comparison to the corresponding literature and the theoretically achievable remanences assuming a composite without porosity. b) Coercivity as function of the filler wt.% in comparison to literature.*

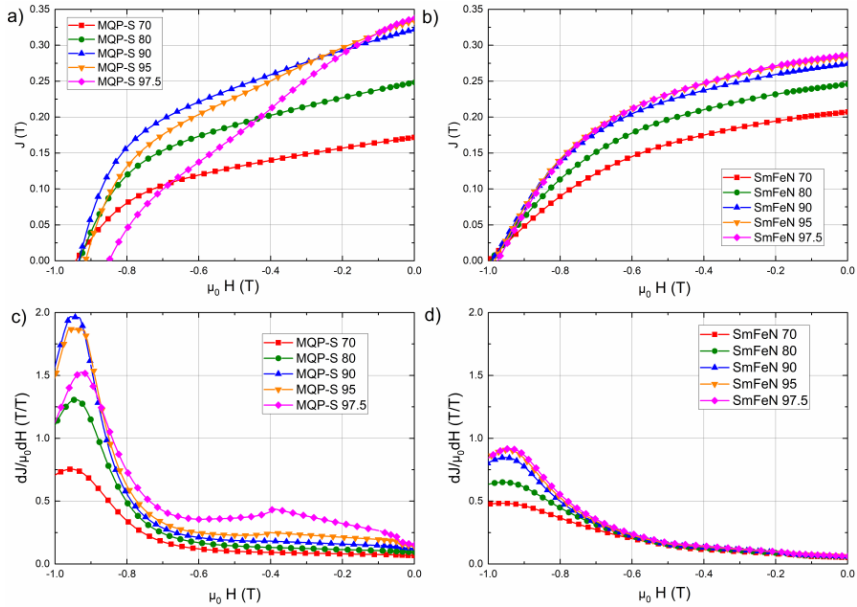

*Figure 12: Second quadrant demagnetization curves J(H) for the MQP-S (a) and SmFeN (b) composites with different wt.%. Magnetic susceptibility for the MQP-S (c) and SmFeN (d) composites with different wt.%.*

Table 5: Literature values of Density, $J_r$, $H_c$ with their corresponding wt.%.

| Reference | Filler | Filler Fraction (Wt. %) | Density (g/cm³) | $J_r$ (mT) /$J_r$ Theo (mT) | $\mu_0 H_c$ (mT) |
|---|---|---|---|---|---|
| SLA [3] Huber | MQP-S | 92 | 4.83 | 388/450 - | 923 |
| FDM [3] Huber | MQP-S | 89 | 3.57 | 344/387 | 918 |
| SLS [3] Huber | MQP-S | 100 | 4.47 | 436/730 | 653 |
| LPBF [12] Mapley 1 | MQP-S | 92 | 3.81 | 362/450 | 923 |
| LPBF [13] Fim | MQP-S | 90 | 4.33 | 385/407 | 906 |
| LPBF [15] Engerroff 2 | SmFeN | 94 | 2.3 – 2.48 | 240-300/621 | 774-806 |
| LPBF [10] Mapley 2 | MQP-S | 96.7 | 3.99 | Non-comparable units/586 | Non-comparable units |
| LPBF: This work | MQP-S | 97.8 | 4.31 | 338/620 | 849 |
| LPBF: This work | SmFeN | 97.6 | 2.28 | 286/771 | 974 |

The remanence $J_r$ and coercivity $H_c$ of the composites with different wt.% in comparison to literature values are presented in Figure 11. The corresponding values can be found in Table 5 together with the density and filler fraction. The corresponding second quadrant demagnetization curves are shown in Figure 12 a) for MQP-S and b) for SmFeN composites and show an increase of the remanence with increasing filler fraction. The remanence of the MQP-S composites increases from 172 mT for the MQP-S 70 sample to a remanence of 338 mT for the MQP-S 97.5 sample. Compared to the theoretical $J_r$ of a fully dense body with the corresponding filler fraction, the discrepancy increases significantly for higher filler fraction, with an up to 45% lower remanence. This divergence between measured and theoretical value can be explained by the significantly increased porosity of the samples at higher filler fractions, which was already observable in agreement with the development of the geometrical density. In comparison with results from literature, the higher density of the SLA sample reported by Huber et al. results in a high remanence of 388 mT [3] and the FDM sample (89 wt.%) by Huber et al. has a similar remanence as the 90 wt.% sample from this work [3]. While these examples show a higher $J_r$ of 86 and 88% of the theoretical limit, the comparison with literature also shows that there is potential for a further improvement of $J_r$ by adjusting the process parameters to achieve higher densities. For the samples produced with the SmFeN filler, a similar behavior can be observed. However, the deviation from the theoretical limit is significantly larger than for MQP-S with only 37 % $J_r$ of the theoretical limit for the SmFeN 97.5 sample. The results shown by Engerroff et al. show similar values with 38-48 % of the theoretical limit [15] and demonstrate that there is a significant need to further optimize the process of LPBF with non-spherical SmFeN fillers.

With approximately 850 mT, the coercivity of the MQP-S 70, 80 and 90 composites are within the range of the initial MQP-S powder with 840 to 940 mT. For higher filler contents of 95 and 97.5 wt.% of MQP-S, it decreases by 2 % and 9 % compared to pure MQP-S or the samples with lower filler fractions, respectively. This observed reduction of the coercivity for high filler fractions can be explained by the increased ability of the magnetic particles to rotate during the demagnetization process. This rotation is enabled by an insufficient bond between the filler and the PA12 matrix, due to only partial wetting during the printing process. Lower volume fractions of the polymer matrix can cause the formation of more porous areas within the sample [13]. These porous areas provide only a few connection points to the MQP-S particles, allowing weaker bonded particles to rotate if an external field is applied. This effect can not only be derived from the decrease of $H_c$ for high filler fractions, but also by an increase of the first derivative of $J(H)$ with increasing filler fraction at applied fields smaller than $H_c$. This change in the magnetic susceptibility shown in Figure 12 c) also indicates that the amount of rotating MQP-S particles increases with higher filler fraction. While the rotation of particles at lower magnetic fields is especially visible in the pronounced peak around 0.4 T for a filler fraction of 97.5 wt.%, the effect is already observable for lower filler fractions of 80 to 95 wt.%. A similar observation regarding the reduction of $H_c$ was also shown by Engerroff et al. for LPBF of milled SmFeN [15]. In their study, the coercivity reduction was up to 14.5 % compared to melt-spun SmFeN flakes and a similar conclusion regarding the influence of the filler fraction was drawn from the results. The energy necessary to rotate the particles and therefore the necessary applied magnetic field are also influenced by the morphology of the filler, as a rotation for highly spherical particles is easier than for more irregular shaped particles. This influence can be seen for the larger platelet shaped SmFeN filler material which cannot rotate as freely.

The coercivity of the SmFeN composites is slightly decreasing from roughly 1000 to 974 mT with increasing filler fraction. In contrast to samples produced with the spherical MQP-S powder, this decrease in $H_c$ is significantly smaller, even for high filler fractions of 97.5 wt.%. Additionally, there is nearly no change in the magnetic susceptibility at low magnetic fields, observable in Figure 12 d), if compared with the previously shown results. This suggests that significantly less rotation of the magnetic particles is present, in agreement with the previously shown density measurements. They indicate an increasing porosity, which should result in the opposite effect on these properties. This agrees well with the expectation of an easier rotation of spherical particles when compared to platelet shaped particles. In comparison with the results shown by Engerroff et al., the coercivity of the printed magnets is at least 20% higher for the values reported in this work. The reason for this increase can be the reduction of $H_c$ for the SmFeN powder during the ball milling process in the study of Engerroff et al., as defects can be introduced in the material. However, also an effect of the morphology cannot be excluded as an influencing factor. When compared to the results for MQP-S, the decrease partially is a product of the increasing porosity and therefore causes worse embedding of the magnetic particles, as well as higher rotatability of the significantly smaller and more spherical ball milled SmFeN particles exhibit in their study.

The use of SmFeN powder with a large platelet shaped morphology shows, that the morphology can be utilized to influence the coercivity development, magnetic susceptibility, and texture in a significant way. This effect needs to be considered for the material choice applied in LPBF, as it might lead to an alternative to the often-considered optimal use of spherical particles.

3.2.5 Mechanical properties:

To further describe the properties of the printed parts, the mechanical properties are discussed for varied filler fraction and the two different magnetic powders. The compression engineering stress strain curves of the pure PA12 and of the composites with different filler fractions are presented in Figure 13 a) for the MQP-S composites and Figure 13 b) for the SmFeN composites. The mean compressive strength at yield $\sigma_y$, mean compressive strength $\sigma_m$, nominal compressive yield strain $\epsilon_y$ and the nominal compressive strain at compressive strength $\epsilon_m$ of five samples per wt.% as function of the filler fraction are shown in Figure 14. Pure PA12 does not fail under compressive condition up to a strain of 60 %. The mechanical stability of both types of composites is significantly reduced and is decreasing with increasing filler fraction. Both $\sigma_y$ and $\sigma_m$ are decreasing with increasing filling fraction. The increase of the compression strength for low filler fractions and a decrease at high filling fractions for polymer composites was also reported by Raja and Kumaravel [9]. The decrease of the compression strength is a result of the increasing porosity which is visible in Figure 10 b) with a rise in porosity for both composites with increasing filling fraction. In addition, the polymer is responsible for the mechanical cohesion of the sample. The literature value from the MQP-S LPBF study fits the trend of the measured values of this work [12].

The mechanical properties of the SmFeN composites are inferior to their MQP-S counterparts with the same wt.%. For example, $\sigma_m$ of the 70 wt.% SmFeN composite is 30 % lower in comparison to the 70 wt.% MQP-S composites. This can be partially explained by the higher porosity of the SmFeN composites. It is also known that the particle morphology has a significant influence on the mechanical properties of a polymer composite [42,43]. It is possible, that the platelets themselves have a lower structural integrity in comparison to the small spheres of MQP-S and can therefore cause failure. Furthermore, the aligned platelet shaped SmFeN particles with sharp edges can be more prone to crack propagation in comparison to the spherical MQPS particles. Further detailed microstructural analysis is necessary to confirm this argumentation.

For the pure PA12 the mean $\sigma_y$ value is 34 % higher in comparison to a similar compression test with pure PA 12 from the literature [44]. The maximal compression strength in the literature study is 60.1 MPa. The maximum strain is 18 %. In accordance with the results from the literature, the compression samples do not break within this strain range. However, the presented data here show a maximum compression strength of 140 MPa at a strain of 60 %. Many factors influence the mechanical behavior of PA 12 processed with LPBF processes like the process parameters and the feedstock material. Both differentiate from the literature which limits the comparability [44]. For example, in the study by Lammens et al. a 50/50 mixture of virgin and recycled powder is used [44] while purely virgin powder was used in the experiments presented here. In addition, the PA-12 powder is from a different supplier (*EOS*) and the LPBF process is performed with a different machine (*P395 EOS*).

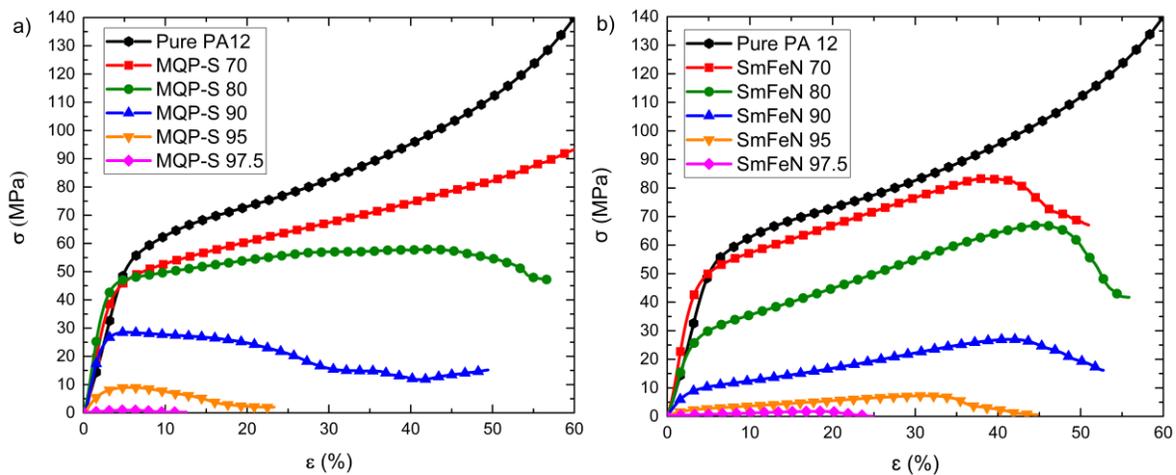

*Figure 13: Compressive engineering stress-strain diagrams of the MQP-S composites a) and the SmFeN composites b) from the position A (see Fig. 2) for composites with different filling fractions.*

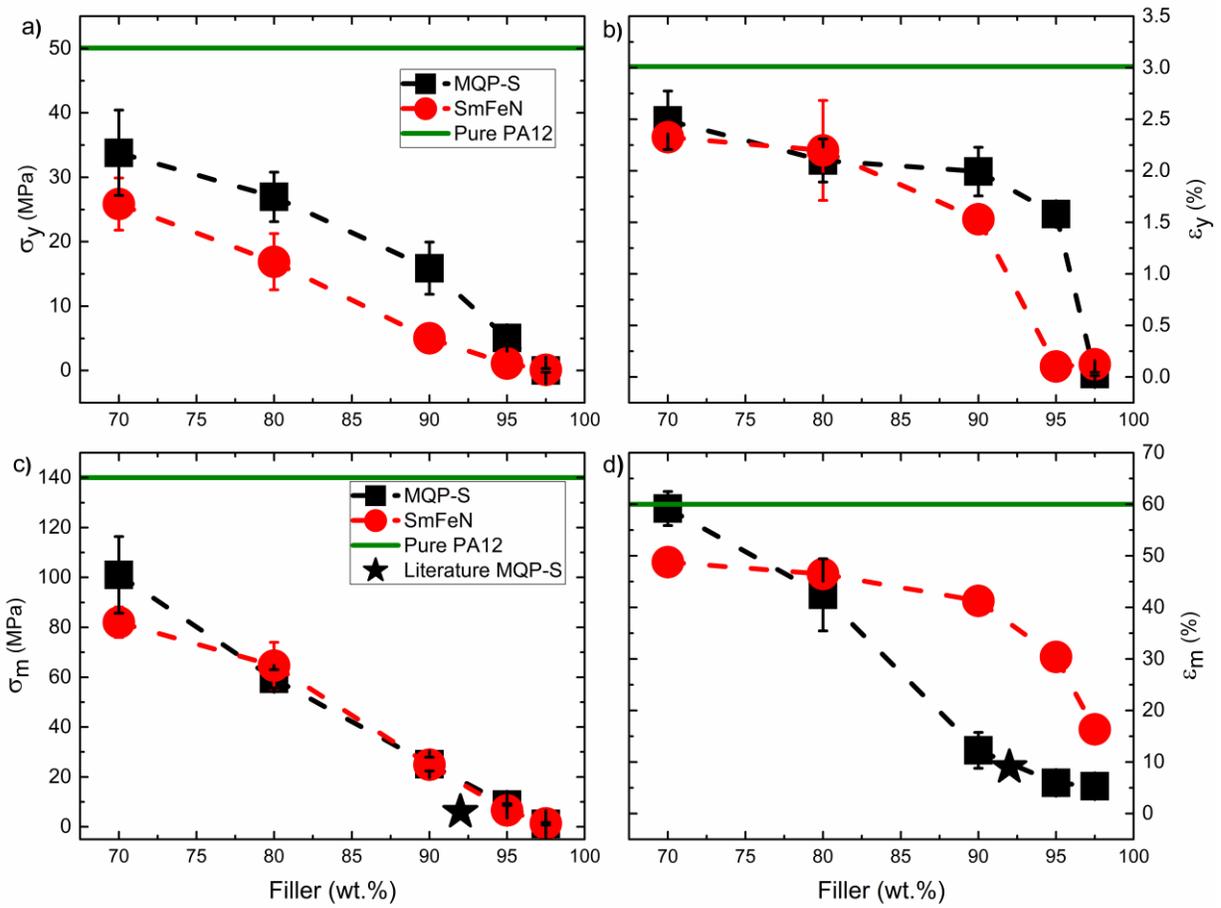

Figure 14: a) The mean compressive strength at yield $\sigma_y$, b) mean compressive strength $\sigma_m$, c) nominal compressive yield strain $\epsilon_y$ and d) the nominal compressive strain at compressive strength $\epsilon_m$ of five samples (Positions A,F,G,H,I) per wt.% as function of the filler fraction.

4. Summary and Outlook

In this work, the microstructural, mechanical and magnetic properties of composite permanent magnets produced by LPBF are investigated depending on process parameters, amount of magnetic filler and powder morphology. It was shown that the laser energy and temperature distribution across the powder bed varies and have a significant impact on the properties of the magnets produced by AM. The geometric density and the mechanical properties decrease significantly with increasing distance from the center of the x-y plane in the powder bed since the amount of thermal energy introduced into the powder decreases. This effect is more pronounced for the application of metal-polymer composites in comparison to the pure PA12 and therefore implies a smaller process window for the production of bonded magnets when they are processed with LPBF. The variation of the magnetic filler fraction shows a difference in the densification of different composites during the AM process. The geometrical density of the MQP-S composites increases with increasing filler fraction while the geometric density of the SmFeN composites first increase from the 70 to 80 wt.% samples and then decreases with increasing filler content. This suggests that the high aspect ratio, platelet-shaped morphology of the SmFeN particles facilitates porosity formation during the LPBF process and that densification is even more impeded for high filling fractions. The mechanical stability of the composites is significantly reduced by higher filling fractions. The remanence of the MQP-S and the SmFeN composites both increases with increasing filler fraction. However, the discrepancy to the theoretical remanence of a fully dense body is up to 45 % lower for MQP-S and 63 % lower for the SmFeN particles. The coercivity of the MQP-S composites significantly decrease with increasing filler fraction. This reduction of the coercivity for high filler fractions can be explained by the increasing probability of insufficient bonded magnetic particles rotating during the demagnetization process. This work extends these results by showing that the filler morphology can influence this change in $H_c$ since rotation is significantly easier for spherical particles than for particles with high aspect ratios like the used SmFeN platelets.

The results indicate that there is a necessity to further optimize the process parameters for these composites, especially for the SmFeN composites, to improve the geometrical density and thus increase the magnetic properties and mechanical stability. For an application the relations between the increasing remanence, slightly decreasing coercivity and significantly decreasing mechanical stability with increasing filler fraction must be considered.

Whilst this study did not show any printed magnets with anisotropic magnetic properties, due to the utilization of isotropic magnetic filler materials, the LPBF study on the microstructure of the SmFeN composites reveals, that platelet-shaped magnetic particles with a high aspect ratio self-align perpendicular towards the build-up direction during the process. This specific effect has a high potential to produce anisotropic magnets without the necessity of an external magnetic field during the process if magnetic anisotropic powder with a platelet-shaped morphology and a magnetic easy axis out of platelet plane are utilized. While the alignment of the platelets slightly deteriorates with higher filler fractions, this effect is a major finding and offers an easy and cost-effective way to produce anisotropic composite magnets in LPBF. Our results illustrate the difficulties and the potential of the utilization of non-spherical particles. Moreover, the *in situ* alignment effect of platelet shaped particles can also be used to tailor other functional properties e.g. electrical conductivity of composites.

Supplementary material

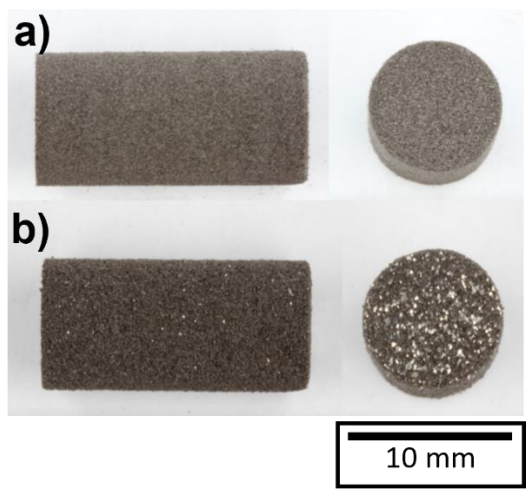

Figure S1: Images of the samples produced by LPBF with a) MQPS and b) SmFeN. Both presented samples are having a filling fraction of 70 wt.%

Declaration of Competing Interest

None.

Acknowledgement

This work was financially supported by the Deutsche Forschungsgemeinschaft (DFG, German Research Foundation), Project ID No. 405553726, TRR 270. The authors thank Daido Steel and Neo Magnequench for providing the powders and Semih Ener for the fruitful discussions.

Author contributions

Kilian Schäfer: Writing – original draft, Conceptualization, Methodology, Investigation Tobias Braun: Writing - original draft, Conceptualization, Methodology, Investigation Stefan Riegg: Conceptualization, Writing - review & editing Jens Musekamp: Investigation - Computed Tomography, Writing - review & editing Oliver Gutfleisch: Funding acquisition, Writing - review & editing, Supervision

# References

[1] O. Gutfleisch, M.A. Willard, E. Brück, C.H. Chen, S.G. Sankar, J.P. Liu, Magnetic materials and devices for the 21st century: stronger, lighter, and more energy efficient, Adv. Mater. Weinheim. 23 (2011) 821–842. https://doi.org/10.1002/adma.201002180.

[2] J. Ormerod, S. Constantinides, Bonded permanent magnets: Current status and future opportunities (invited), J. Appl. Phys. 81 (1997) 4816–4820. https://doi.org/10.1063/1.365471.

[3] C. Huber, G. Mitteramskogler, M. Goertler, I. Teliban, M. Groenefeld, D. Suess, Additive Manufactured Polymer-Bonded Isotropic NdFeB Magnets by Stereolithography and Their Comparison to Fused Filament Fabricated and Selective Laser Sintered Magnets, Materials 13 (2020) 1916. https://doi.org/10.3390/ma13081916.

[4] L. Li, A. Tirado, I.C. Nlebedim, O. Rios, B. Post, V. Kunc, R.R. Lowden, E. Lara-Curzio, R. Fredette, J. Ormerod, T.A. Lograsso, M.P. Paranthaman, Big Area Additive Manufacturing of High Performance Bonded NdFeB Magnets, Sci. Rep. 6 (2016) 36212. https://doi.org/10.1038/srep36212.

[5] L. Li, B. Post, V. Kunc, A.M. Elliott, M.P. Paranthaman, Additive manufacturing of near-net-shape bonded magnets: Prospects and challenges, Scripta Mater. 135 (2017) 100–104. https://doi.org/10.1016/j.scriptamat.2016.12.035.

[6] C. Huber, C. Abert, F. Bruckner, M. Groenefeld, S. Schuschnigg, I. Teliban, C. Vogler, G. Wautischer, R. Windl, D. Suess, 3D Printing of Polymer-Bonded Rare-Earth Magnets With a Variable Magnetic Compound Fraction for a Predefined Stray Field, Sci. Rep. 7 (2017) 9419. https://doi.org/10.1038/s41598-017-09864-0.

[7] E. Peng, X. Wei, T.S. Herng, U. Garbe, D. Yu, J. Ding, Ferrite-based soft and hard magnetic structures by extrusion free-forming, RSC Adv. 7 (2017) 27128–27138. https://doi.org/10.1039/C7RA03251J.

[8] N. Urban, A. Meyer, V. Keller, J. Franke, Contribution of Additive Manufacturing of Rare Earth Material to the Increase in Performance and Resource Efficiency of Permanent Magnets, AMM 882 (2018) 135–141. https://doi.org/10.4028/www.scientific.net/AMM.882.135.

[9] V.L. Raja and A. Kumaravel, Raja VL, Kumaravel A. Studies on Physical and Mechanical Properties of Silica Fume-Filled Nylon 66 Polymer Composites for Mechanical Components. Polym. Polym. Compos. (2015):427-434. doi:10.1177/096739111502300608

[10] M. Mapley, S.D. Gregory, J.P. Pauls, G. Tansley, A. Busch, Influence of Powder Loading Fraction on Properties of Bonded Permanent Magnets Prepared By Selective Laser Sintering, 3D Print. Addit. Manuf. 8.3 (2021): 168-175. https://doi.org/10.1089/3dp.2020.0297

[11] S. Yuan, F. Shen, C.K. Chua, K. Zhou, Polymeric composites for powder-based additive manufacturing: Materials and applications, Prog. Polym. Sci. 91 (2019) 141–168. https://doi.org/10.1016/j.progpolymsci.2018.11.001.

[12] M. Mapley, J.P. Pauls, G. Tansley, A. Busch, S.D. Gregory, Selective laser sintering of bonded magnets from flake and spherical powders, Scripta Mater. 172 (2019) 154–158. https://doi.org/10.1016/j.scriptamat.2019.07.029.

[13] R. Fim, A.A. Mascheroni, L.F. Antunes, J. Engerroff, C.H. Ahrens, P. Wendhausen, Increasing packing density of Additively Manufactured Nd-Fe-B bonded magnets, Addit. Manuf. 35 (2020) 101353. https://doi.org/10.1016/j.addma.2020.101353.

[14] T. Fiegl, M. Franke, C. Körner, Impact of build envelope on the properties of additive manufactured parts from AlSi10Mg, Opt. Laser Technol. 111 (2019) 51–57. https://doi.org/10.1016/j.optlastec.2018.08.050.

[15] J. Engerroff, A.B. Baldissera, M.D. Magalhães, P.H. Lamarão, P. Wendhausen, C.H. Ahrens, J.M. Mascheroni, Additive manufacturing of Sm-Fe-N magnets, J. Rare Earths 37 (2019) 1078–1082. https://doi.org/10.1016/j.jre.2019.04.012.

[16] K. Gandha, L. Li, I.C. Nlebedim, B.K. Post, V. Kunc, B.C. Sales, J. Bell, M.P. Paranthaman, Additive manufacturing of anisotropic hybrid NdFeB-SmFeN nylon composite bonded magnets, J. Magn. Magn. Mater. 467 (2018) 8–13. https://doi.org/10.1016/j.jmmm.2018.07.021.

[17] Neo Magnequench, MQP-S data sheet.

[18] J. Coey, P. Stamenov, S.B. Porter, M. Venkatesan, R. Zhang, T. Iriyama, Sm-Fe-N revisited; remanence enhancement in melt-spun Nitroquench material, J. Magn. Magn. Mater. 480 (2019) 186–192. https://doi.org/10.1016/j.jmmm.2019.02.076.

[19] O Gutfleisch, Controlling the properties of high energy density permanent magnetic materials by different rocessing routes. J. Phys. D. Appl. Phys. 33.17 (2000): R157. https://doi.org/10.1088/0022-3727/33/17/201

[20] Sintratec AG, PA 12 data sheet.

[21] Daido Steel, SmFeN data sheet.

[22] DIN EN ISO 604. Plastics: Determination of pressure properties. 2003.

[23] A. Aharoni, Demagnetizing factors for rectangular ferromagnetic prisms, J. Appl. Phys. 83 (1998) 3432–3434. https://doi.org/10.1063/1.367113

[24] J. Schindelin, I. Arganda-Carreras, E. Frise, V. Kaynig, M. Longair, T. Pietzsch, S. Preibisch, C. Rueden, S. Saalfeld, B. Schmid, J.-Y. Tinevez, D.J. White, V. Hartenstein, K. Eliceiri, P. Tomancak, A. Cardona, Fiji: an open-source platform for biological-image analysis, Nat. Methods 9 (2012) 676–682. https://doi.org/10.1038/nmeth.2019.

[25] Z. Püspöki, M. Storath, D. Sage, M. Unser, Transforms and operators for directional bioimage analysis: a survey, Focus on bio-image informatics (2016) 69–93. DOI: 10.1007/978-3-319-28549-8_3

[26] I. Gibson, D. Rosen, B. Stucker, M. Khorasani, Additive Manufacturing Technologies, Springer International Publishing, Cham, 2021. https://doi.org/10.1007/978-3-030-56127-7

[27] J.D. Nisi, F. Pozzi, P. Folgarait, G. Ceselin, M. Ronci, Precipitation hardening stainless steel produced by powder bed fusion: influence of positioning and heat treatment, Procedia Structural Integrity 24 (2019) 541–558. https://doi.org/10.1016/j.prostr.2020.02.048.

[28] A. Unkovskiy, P.H.-B. Bui, C. Schille, J. Geis-Gerstorfer, F. Huettig, S. Spintzyk, Objects build orientation, positioning, and curing influence dimensional accuracy and flexural properties of stereolithographically printed resin, Dent. Mater. 34 (2018) e324-e333. https://doi.org/10.1016/j.dental.2018.09.011.

[29] M. Schmid, A. Amado, K. Wegener, Materials perspective of polymers for additive manufacturing with selective laser sintering, J. Mater. Res. 29 (2014) 1824–1832. https://doi.org/10.1557/jmr.2014.138.

[30] P.-M. Yang, Y.-L. Lo, Y.-H. Chang, Laser galvanometric scanning system for improved average power uniformity and larger scanning area, Appl. Opt. 55 (2016) 5001–5007. https://doi.org/10.1364/AO.55.005001.

[31] H. Zarringhalam, C. Majewski, N. Hopkinson, Degree of particle melt in Nylon-12 selective laser sintered parts. Rapid Prototyp. J. (2009). Vol. 15 No. 2, pp. 126-132. https://doi.org/10.1108/13552540910943423

[32] S. Dupin, O. Lame, C. Barrès, J.-Y. Charmeau, Microstructural origin of physical and mechanical properties of polyamide 12 processed by laser sintering, Eur. Polym.J. 48 (2012) 1611–1621. https://doi.org/10.1016/j.eurpolymj.2012.06.007.

[33] G. Liao, Z. Li, Y. Cheng, D. Xu, D. Zhu, S. Jiang, J. Guo, X. Chen, G. Xu, Y. Zhu, Properties of oriented carbon fiber/polyamide 12 composite parts fabricated by fused deposition modeling, Mater. Des. 139 (2018) 283–292. https://doi.org/10.1016/j.matdes.2017.11.027.

[34] D. Zhu, Y. Ren, G. Liao, S. Jiang, F. Liu, J. Guo, G. Xu, Thermal and mechanical properties of polyamide 12/graphene nanoplatelets nanocomposites and parts fabricated by fused deposition modeling, J. Appl. Polym. Sci. 134 (2017) 45332. https://doi.org/10.1002/app.45332.

[35] R.M. Erb, R. Libanori, N. Rothfuchs, A.R. Studart, Composites reinforced in three dimensions by using low magnetic fields, Science 335 (2012) 199–204. DOI: 10.1126/science.1210822

[36] C.C. Furnas, Grading aggregates-I.-Mathematical relations for beds of broken solids of maximum density, Industrial & Engineering Chemistry 23 (1931) 1052–1058.

[37] E.C. Abdullah, D. Geldart, The use of bulk density measurements as flowability indicators, Powder Technol. 102 (1999) 151–165. https://doi.org/10.1016/S0032-5910(98)00208-3

[38] J.-P. Kruth, G. Levy, F. Klocke, T. Childs, Consolidation phenomena in laser and powder-bed based layered manufacturing, CIRP Annals 56 (2007) 730–759. https://doi.org/10.1016/j.cirp.2007.10.004.

[39] K. Wudy, D. Drummer, Aging effects of polyamide 12 in selective laser sintering: Molecular weight distribution and thermal properties, Addit. Manuf. 25 (2019) 1–9. https://doi.org/10.1016/j.addma.2018.11.007.

[40] Saini, A.V. Shenoy, V.M. Nadkarni, Melt rheology of highly loaded ferrite-filled polymer composites, Polymer composites 7 (1986) 193–200. https://doi.org/10.1002/pc.750070402

[41] S. Yerazunis, J.W. Bartlett, A.H. Nissan, Packing of binary mixtures of spheres and irregular particles, Nature 195 (1962) 33–35. https://doi.org/10.1038/195033a0

[42] Z. Lin, Y. Liu, S. Raghavan, K. Moon, S.K. Sitaraman, C. Wong, Magnetic alignment of hexagonal boron nitride platelets in polymer matrix: toward high performance anisotropic polymer composites for electronic encapsulation, ACS Appl. Mater. Interfaces 5 (2013) 7633–7640. https://doi.org/10.1021/am401939z.

[43] J.J. Martin, B.E. Fiore, R.M. Erb, Designing bioinspired composite reinforcement architectures via 3D magnetic printing, Nat. Commun. 6 (2015) 8641. https://doi.org/10.1038/ncomms9641.

[44] N. Lammens, M. Kersemans, I. de Baere, W. van Paepegem, On the visco-elasto-plastic response of additively manufactured polyamide-12 (PA-12) through selective laser sintering, Polymer Test. 57 (2017) 149–155. https://doi.org/10.1016/j.polymertesting.2016.11.032.